\documentclass{sigchi}

\usepackage{balance}       
\usepackage{graphics}      
\usepackage[T1]{fontenc}   
\usepackage{txfonts}
\usepackage{mathptmx}
\usepackage[pdflang={en-US},pdftex]{hyperref}
\usepackage{color}
\usepackage{booktabs}
\usepackage{textcomp}

\usepackage{microtype}        
\usepackage{ccicons}          
\usepackage[small,bf,BF]{subfigure}
\usepackage{multirow}
\usepackage{todonotes}
\usepackage{graphicx}
\usepackage{mwe}
\usepackage{enumitem}
\usepackage{xspace}
\usepackage{xstring}
\newcommand*\rot{\rotatebox{90}} 

\usepackage{ragged2e}

\makeatletter
\renewcommand\p@subfigure{}
\makeatother

\def\plaintitle{GLADAS: Gesture Learning\texorpdfstring{\\for Advanced Driver Assistance Systems}{} }

\def\plainkeywords{Self-Driving Cars; Hand Gestures; Pedestrians; Communication}

\makeatletter
\def\url@leostyle{%
  \@ifundefined{selectfont}{
    \def\UrlFont{\sf}
  }{
    \def\UrlFont{\small\bf\ttfamily}
  }}
\makeatother
\def\pprw{8.5in}
\def\pprh{11in}

\def\emptyauthor{}

\definecolor{linkColor}{RGB}{6,125,233}
\hypersetup{%
  pdftitle={\plaintitle},
  pdfauthor={\emptyauthor},
  pdfkeywords={\plainkeywords},
  pdfdisplaydoctitle=true, 
  bookmarksnumbered,
  pdfstartview={FitH},
  colorlinks,
  citecolor=green,
  filecolor=black,
  linkcolor=red,
  urlcolor=linkColor,
  breaklinks=true,
  hypertexnames=false
}

\begin{document}

\newcommand{\gladas}[0]{GLADAS\xspace}

\title{\plaintitle \vspace*{20pt}}

\numberofauthors{3}
\author{%
  \alignauthor{Ethan Shaotran\thanks{\textit{This work was done while Ethan was a visiting student at Harvard under the Research Science Institute (RSI) summer STEM research program for gifted and talented high school students.}}\\
    \affaddr{Harvard University}\\
    \email{\tt shaotran@mit.edu}}\\
  \alignauthor{Jonathan J. Cruz\\
    \affaddr{Harvard University}\\
    \email{\tt jcruz@g.harvard.edu}\\
    \vspace{5pt}
    \mbox{
        \hspace{-10pt}
        \url{https://github.com/harvard-edge/gladas}
    }}
  \alignauthor{Vijay Janapa Reddi\\
    \affaddr{Harvard University}\\
    \email{\tt vj@eecs.harvard.edu}}\\
}

\maketitle


\textbf{
\hspace{5pt} \textit{Abstract}---Human-computer interaction (HCI) is crucial for the safety of lives as autonomous vehicles (AVs) become commonplace. Yet, little effort has been put toward ensuring that AVs understand humans on the road. In this paper, we present \gladas, a simulator-based research platform designed to teach AVs to understand pedestrian hand gestures. \gladas supports the training, testing, and validation of deep learning-based self-driving car gesture recognition systems. We focus on gestures as they are a primordial (i.e, natural and common) way to interact with cars. To the best of our knowledge, \gladas is the first system of its kind designed to provide an infrastructure for further research into human-AV interaction. We also develop a hand gesture recognition algorithm for self-driving cars, using \gladas to evaluate its performance. Our results show that an AV understands human gestures 85.91\% of the time, reinforcing the need for further research into human-AV interaction.
}

\ccsdesc[500]{Human-centered computing~Gestural input}
\ccsdesc[500]{Applied computing~Transportation}
\ccsdesc[100]{Computer systems organization~Robotic autonomy}

\keywords{\plainkeywords}


\section{Introduction}
\label{sec:intro}


The advent of Self-Driving Car (SDC) technology is anticipated to improve transportation costs and congestion, reduce traffic accidents, and even mitigate climate change \cite{arbib_seba_2017}. Realizing these benefits will require autonomous vehicles to feature strong human-computer interaction skills, among other factors \cite{Rasouli_Tsotsos_2018, Lahijanian_Kwiatkowska_2016}. However, little progress has been made towards human-AV interaction in road driving scenarios. Situations in which an SDC and pedestrian must decide who should cross first at an intersection, for example, will not be possible if the car is not able to properly respond to such interactions.

Small-scale research efforts have been made, especially in car-to-pedestrian communication. Vinkhuyzen and Cefkin \cite{Vinkhuyzen_Cefkin_2016}, and Habibovic et al. \cite{Habibovic_Lundgren_Andersson_Klingegard_Lagstrom_Sirkka_Fagerlonn_Edgren_Fredriksson_Krupenia_etal._2018}, developed a light strip interface to indicate that pedestrians could cross. Mahadevan et al. \cite{Mahadevan_Somanath_Sharlin_2018} furthered these efforts by additionally attaching an LCD display with a face on it, which indicated the vehicle's awareness of a pedestrian. Matthews et al. \cite{Matthews_Chowdhary_Kieson_2017} used several devices---an LED strip, an LED word display panel, a speaker, and a strobe light---to communicate with pedestrians. All four studies found that pedestrians reacted positively to the addition of these intent-conveying interfaces.

We instead focus on pedestrian-to-car communication, in which the SDC must understand the pedestrian's intent. Vinkhuyzen and Cefkin \cite{Vinkhuyzen_Cefkin_2016} find hand gestures to be a customary practice in negotiating the right of way---for example, a pedestrian may give up his/her right of way by using a hand gesture to signal a car to proceed first at an intersection. We propose \gladas, an open-source research platform designed to support the testing and benchmarking of gesture recognition algorithms of self-driving cars.

We develop \gladas to feature a virtual simulation with common Car-Pedestrian Interaction (CPI) scenarios. Each pedestrian is animated with five different hand gestures commonly used in roadway situations. To demonstrate the abilities of \gladas to support Gesture Learning (GL) tests, we create a simple pedestrian hand gesture recognition algorithm, designed to be replaceable with others, that models one such method SDCs may use for gesture recognition. The algorithm is composed of an efficient two-model architecture, which identifies any potential pedestrians and classifies their gesture. Finally, \gladas tests the algorithm in four different CPI scenarios for a total of 28,000 times.

Simulation results show that an SDC classifies human hand gestures with an associated F1 score of 85.91\%. In the context of an SDC, these results are poor, as an SDC crashing potentially 14.09\% of the time is far from ideal. These results, the first of their kind, set a historical baseline for future work to improve on, and highlight the growing need for more extensive research on human-AV interaction. 

Future autonomous vehicles will be required to have strong social interaction capabilities in order to handle the many encounters with pedestrians, particularly in busy urban environments. Our work also acts as an impetus for future benchmark systems of Advanced Driver Assistance Systems (ADAS) designed to test their performance on pedestrian-to-car communication tests, similar to today's seat belt safety tests. Our bottom line is that we  need a more systematic and quantitative-driven approach for testing and analysis of autonomous vehicles.

In summary, we make the following contributions:

\begin{itemize}
	\item We develop the first autonomous driving simulator for focused research on pedestrian-to-car communication and Gesture Learning (GL).
	\item We describe a new end-to-end methodology for training self-driving cars to understand and predict pedestrian behavior to enable safer roads in the future.
	\item We quantitatively motivate the need for more research on human-AV interaction, without which the successful deployment of SDCs may be limited to a small handful of use cases (e.g. driving on highways versus driving through busy city streets with humans).
\end{itemize}

In the remainder of the paper, we debrief related work on the topic (Section~\ref{sec:related}), and then describe the various parts of the \gladas (Section~\ref{sec:gladas},\ref{sec:algorithm}) framework. Then, we use \gladas to measure the performance of a gesture recognition algorithm (Section~\ref{sec:expsetup}), and discuss the findings and their implications on the future of human-computer interaction for autonomous vehicles (Section~\ref{sec:results}). Finally, we conclude the paper (Section~\ref{sec:conc}).

\section{Related Works}
\label{sec:related}

Research on SDC-pedestrian communication can be split into two distinct categories. The majority focus of related work is on the conveyance of a message from an SDC to a pedestrian, which we call \textit{car-to-pedestrian communication} (such as \cite{Vinkhuyzen_Cefkin_2016,Habibovic_Lundgren_Andersson_Klingegard_Lagstrom_Sirkka_Fagerlonn_Edgren_Fredriksson_Krupenia_etal._2018, Mahadevan_Somanath_Sharlin_2018, Matthews_Chowdhary_Kieson_2017}). \textit{Pedestrian-to-car communication}, the reception of a pedestrian's intent by an SDC, has rarely been researched, and is the basis of our work.

Both Uber and Waymo intend to have SDCs adapt to the world as-is, without having to change innate human behavior \cite{insurancejournal_2018}. In the case of Pedestrian-to-car communication, this means that research must focus on the car-side (training SDCs to understand pedestrians' messages well), rather than on the pedestrian-side (training pedestrians to deliver messages to SDCs well). Several methods have been used to do so. Rasouli et al. \cite{Rasouli_Kotseruba_Tsotsos_2018} demonstrated prediction of pedestrian behavior based on head orientation. Kim et al. \cite{Kim_Guy_Liu_WLau_Lin_Manocha_2014} developed a model for predicting pedestrian trajectories based on their current trajectory and kinematics. Schneemann and Heinemann \cite{Schneemann_Heinemann_2016} explore the surrounding street structure as a potential indicator of future pedestrian behavior.

These methods attempt to solve the problem by looking at a pedestrian's implicit communication, in which messages are inferred by an observer based on a human's actions. For example, a pedestrian who is about to cross a roadway may glance in a certain direction or physically step into the road. In contrast, the basis of our work is explicit communication, in which the pedestrian directly conveys a message to a car.

The social relevancy of explicit communication in roadway scenarios is widely noted. Rasouli and Tsotsos \cite{Rasouli_Kotseruba_Tsotsos_2017} find that hand gestures and nodding are prominent forms of explicit communication that pedestrians use to directly communicate with cars. Furthermore, Gupta et al. \cite{Gupta_Vasardani_Winter_2016} have researched common commands used in human-computer interaction in roadway scenarios (``Stop'', ``Turn Right'', etc.).   

Only a limited number of studies have been conducted on recognizing human hand gestures in a roadway setting. Tao and Ben \cite{Tao_Ben} developed an accelerometer, which is placed in the surrounding intersection, to classify Chinese policepeople's gestures and mirror the command in the above traffic lights. Guo et al. \cite{Guo_Tang_Zhu_2015} employed statistical techniques and the nearest neighbor classifier for recognizing gestures in still, staged images of Chinese policepeople taken by normal cameras. Their results indicated accuracies between 60\% to 100\% for each hand gesture class. 

It is evident that human-AV interaction is a poorly-explored problem, especially in the context of understanding hand gestures, requiring more research. A comprehensive effort to classify hand gestures for all pedestrians, not just policepeople, in roadway scenarios is needed. Additionally, the gesture recognition should be trained from the first-person point of view of an SDC---more in line with the manner in which real-world SDCs perceive their environments. \gladas, which tests and validates deep learning-based gesture recognition algorithms that classify gestures from real-time video streams of SDC sensors on the roadway, seeks to fill these gaps.

\section{GLADAS Methodology}
\label{sec:gladas}

\subsection{Overview}
\begin{figure}
\centering
  \includegraphics[width=0.9\columnwidth]{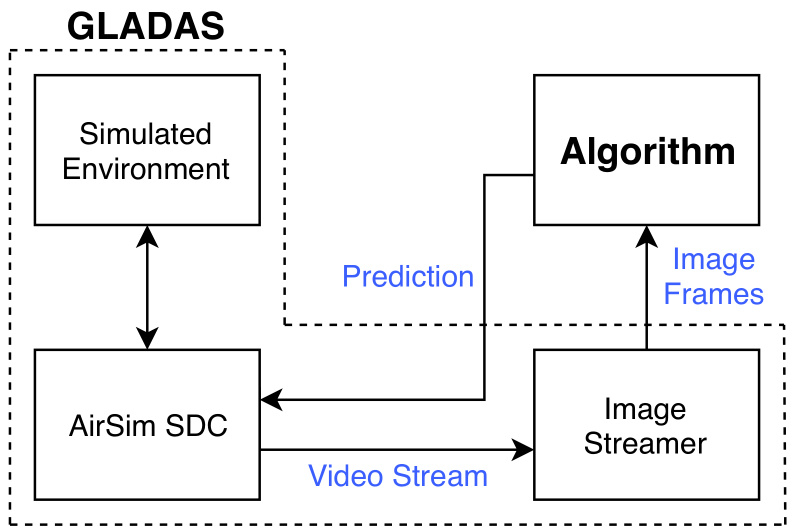}
  \caption{The GLADAS framework features three main parts: a simulated environment, AirSim SDC, and image streamer. An algorithm can then be "plugged in" and benchmarked. }~\label{fig:GLADSFramework}
\end{figure}

We conduct our research in a controlled, virtual simulation with pedestrians, capable of making hand gestures. Our rationale for using a simulator is simple: conducting real-world tests with SDCs and real people is not an option. A simulated SDC streams real-time images to our hand gesture recognition algorithm, which controls the movement of the car.

We set up our simulation environment with three principles in mind: (1) safety of the public, (2) practicality, and (3) flexibility to experiment with parameters.

\subsubsection{Safety} Our simulation allows us to rapidly and repeatedly test the car in a given scenario, with pre-labeled testing data and minimized risk to the general public \cite{OKelly_Sinha_Namkoong_Duchi_Tedrake_2018}. Real-life testing would require us to drive to different intersections and set up an SDC-pedestrian interaction at each one. This is impractical and potentially hazardous---for example, an SDC could act on an incorrectly-identified gesture and crash into the pedestrian. A simulated world and self-driving car provide us with a stable and reliable platform, in which the car's mistakes have no real-world consequences. 

\subsubsection{Realism} Research has shown that simulated SDC data can be successfully used as training data for real-world SDCs \cite{Johnson-Roberson_Barto_Mehta_Sridhar_Rosaen_Vasudevan_2016}. SDC simulators like the CARLA platform \cite{Dosovitskiy_Ros_Codevilla_Lopez_Koltun_2017} allow developers to train their AV algorithms faster and more efficiently.

\subsubsection{Ease of Experimentation} We are able to adjust several factors such as the objects in the surrounding environment, the layout of the roads, and the position of the pedestrian. This provides a wide range of driving situations that can be simulated within \gladas. Additionally, an exact scenario can be constructed within the simulation, without the need to consider any practical constraints associated with real-world testing---ie. having to find a perfect pre-existing area, closing it to the public, and finding volunteers to participate. While \gladas can simulate exact scenarios for testing, it can also simulate completely random scenarios, allowing for rigorous testing of an SDC's performance and reliability.

\subsection{\gladas Architecture}
To achieve the aforementioned objectives, \gladas is composed from three main parts as shown in Figure~\ref{fig:GLADSFramework}: The Unreal Simulated Environment, AirSim, and Image Streamer. We elaborate on the three components of \gladas in this section, and dedicate the next section entirely to a detailed description of the test hand gesture recognition algorithm.

\subsubsection{Unreal Simulated Environment}

\begin{figure}%
\centering
\subfigure[Street view.]{%
\label{fig:SV1}
\includegraphics[width=\columnwidth]{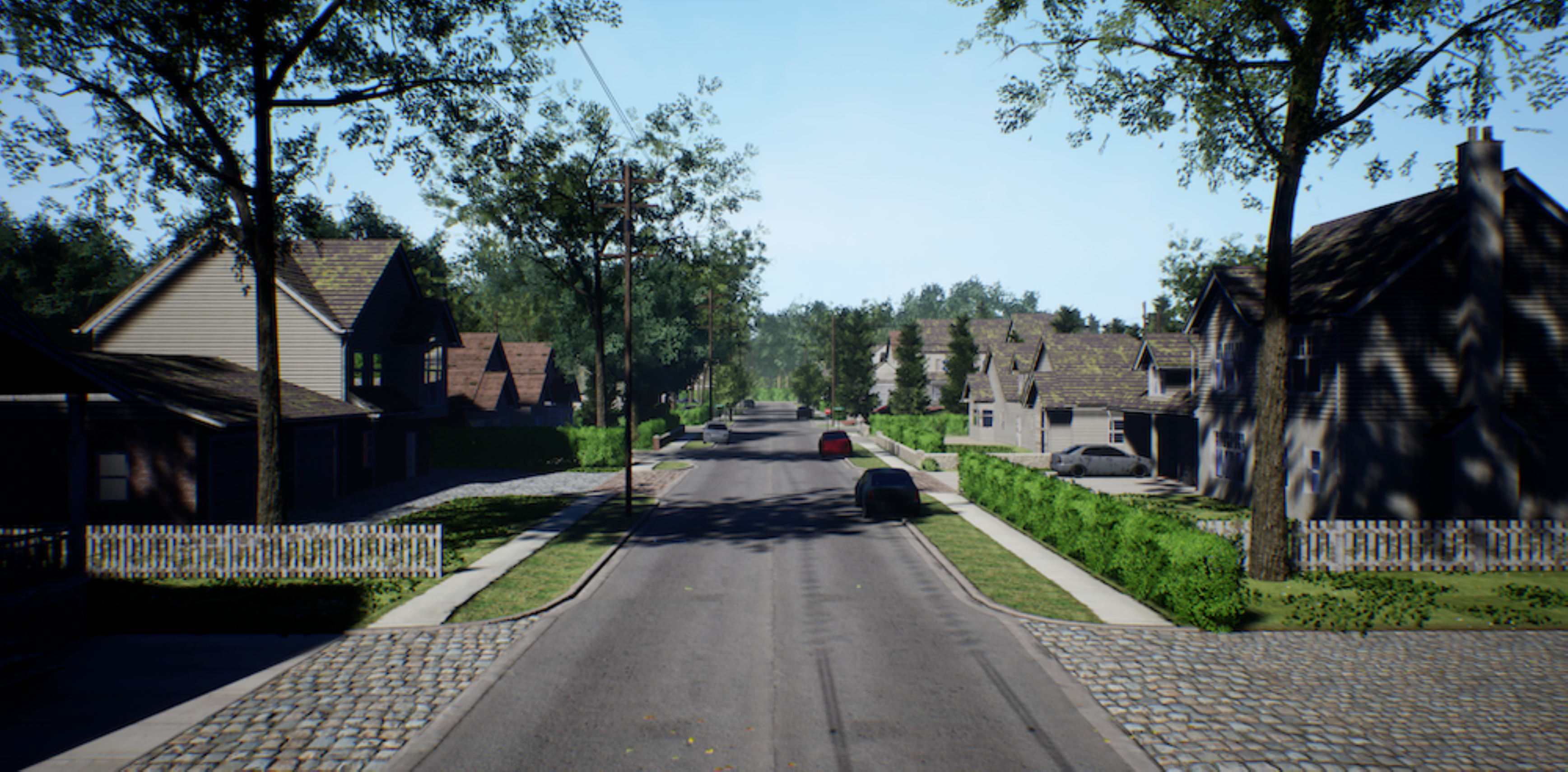}}%
\qquad
\centering
\subfigure[View from inside the self-driving car.]{%
\label{fig:SV2}
\includegraphics[width=\columnwidth]{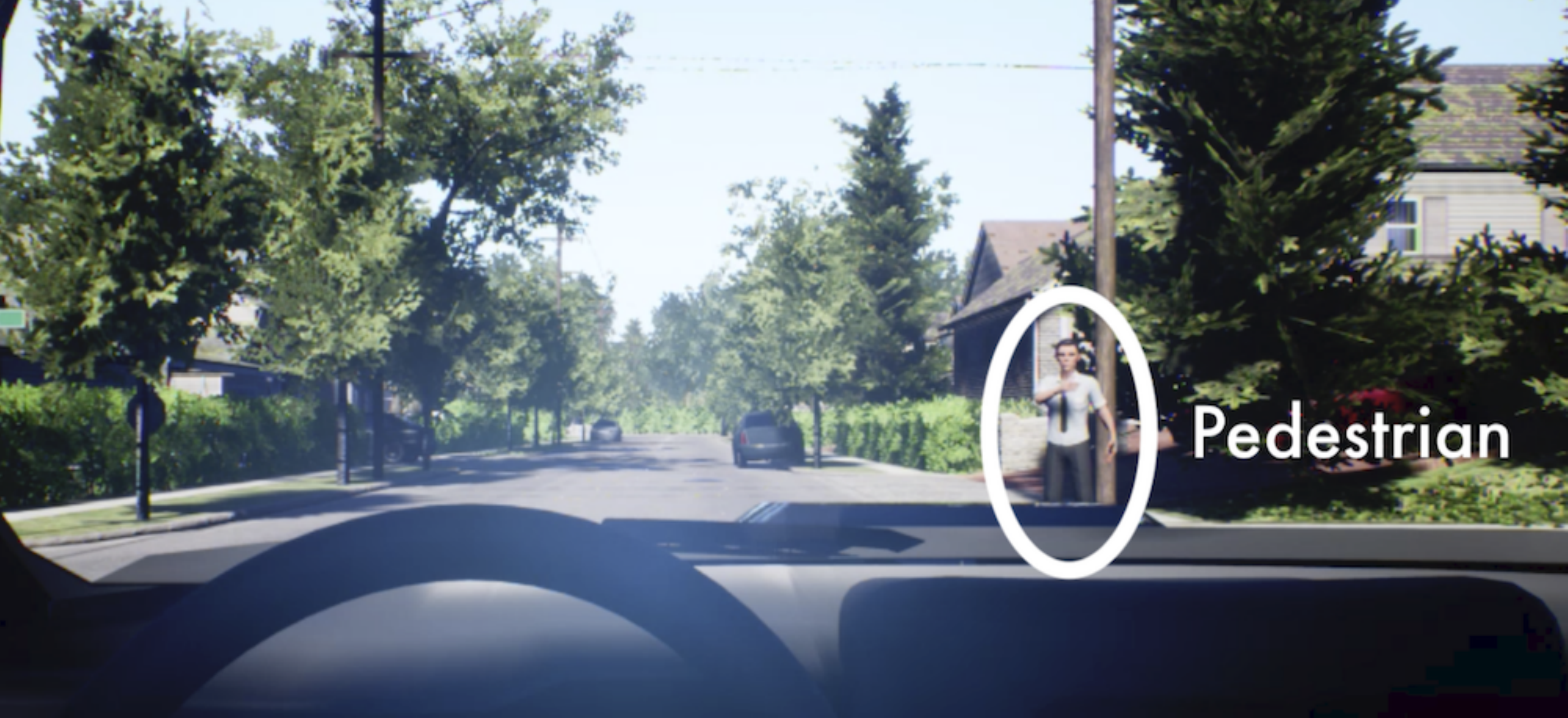}}%
\caption{3D environment used to simulate real-world scenarios.}~\label{fig:SimulatorView}
\end{figure}

We build our simulation in Unreal Engine 4, a 3D game engine commonly used for SDC simulations \cite{Rosique_Navarro_Fernandez_Padilla_2019}. To make our simulation as representative of real-world driving as possible and emulate scenarios an SDC might visually experience in reality, we add assets such as houses, parked cars, roads, road signs, and vegetation, shown in ~\ref{fig:SV1}. The view from within the car is shown in ~\ref{fig:SV2}, with a human waving at the car.

\begin{figure*}
  \centering
  \includegraphics[width=1.6\columnwidth]{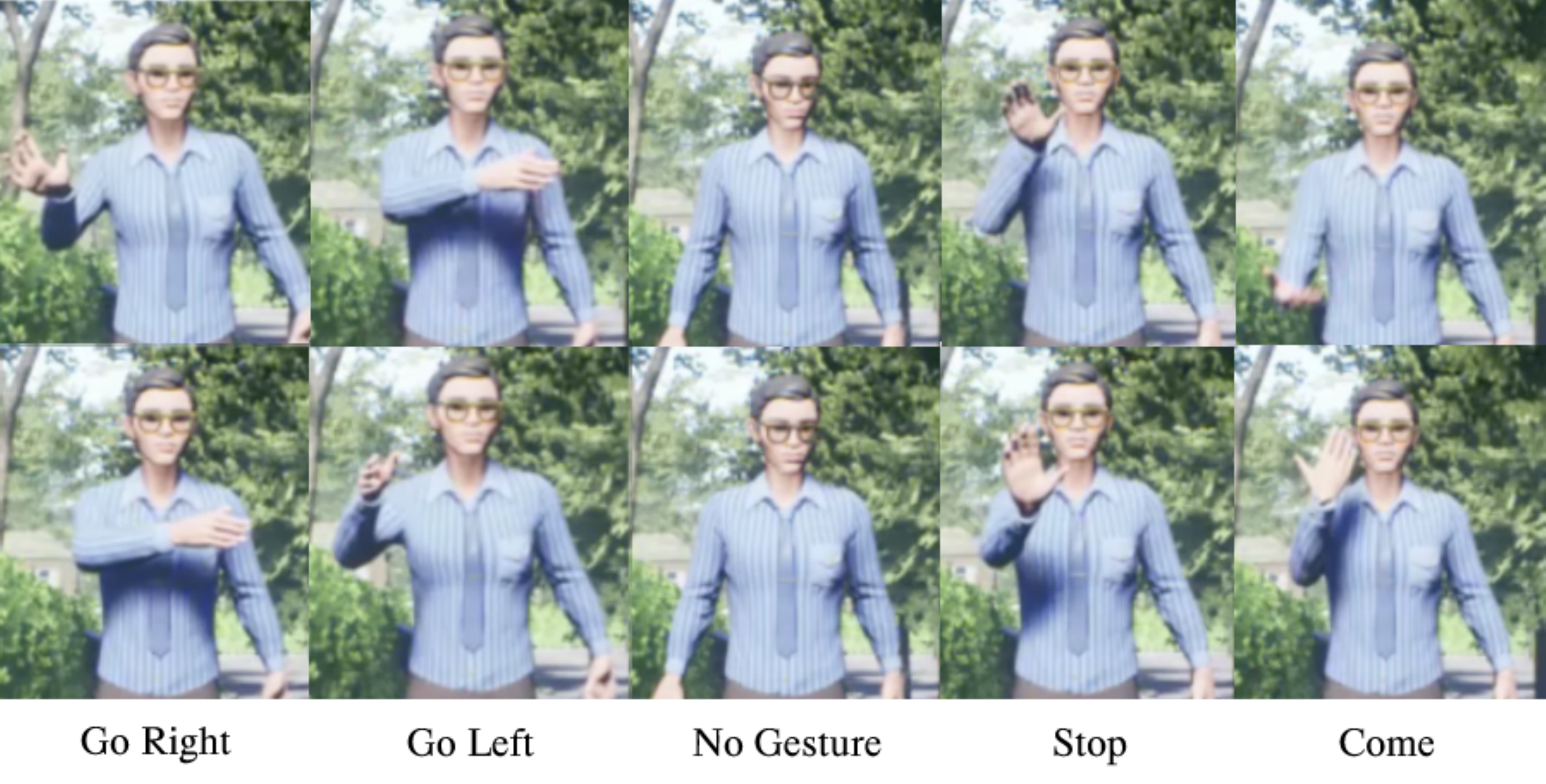}
  \caption{Each column shows representative frames from one of the gestures used to communicate between pedestrians and cars in our work. The top frames represent the beginning of each gesture, and the bottom frames represent the end of each gesture.}~\label{fig:GestureExamples}
\end{figure*}

In order to test our gesture recognition algorithm, we add clothed pedestrians with animated gestures. The commands associated with each gesture, as suggested by Gupta et al. \cite{Gupta_Vasardani_Winter_2016}, are the four most commonly-used commands in road driving scenarios: ``Go Forward'', ``Stop'', ``Go Right'', and ``Go Left''. We also add a base class ``No Gesture'' for a total of five classes. Representative frames from each of the four gestures are shown in Figure~\ref{fig:GestureExamples}. 

Hand gestures can vary across different geographical locations and cultures (for example, the pullover command is a right arm straight up in Germany and both arms straight up in India \cite{Gupta_Vasardani_Winter_2016}), making the creation of an overarching, robust model difficult. We stage our research as an initial study, experimenting with hand gesture recognition for 5 gestures, with the intent that future work can use this methodology and re-evaluate SDCs with other gesture variations that might be used to convey the same commands. We have designed \gladas to be easily extendable to other gestures.

\subsubsection{AirSim}

We use the AirSim~\cite{Shah_Dey_Lovett_Kapoor_2017} Unreal plugin for simulation of SDCs in the environment. AirSim allows us to add LIDAR, depth, and camera sensors to our car, allowing for multi-modal, real-time perception of the car's environment. These three sensors are commonly found on real-world SDCs \cite{DeSilva_Roche_Kondoz_2017}, making our simulated SDC's sensor suite representative of real-life SDCs.

The RGB camera takes a picture with three color channels, displaying colors in any combination of the colors red, green, and blue. LIDAR sensors use lasers to measure distances, creating a three-dimensional point cloud of the surrounding environment. Depth Camera sensors measure distances between the car and objects in front of it (the road, other cars, people, etc.) with infra-red projectors and cameras, in order to create a two-dimensional visualization of depth values. 

We solely utilize data from the RGB camera for gesture recognition. Industry-developed SDCs mainly use LIDAR for mapping the surrounding infrastructure, landscape, and foliage \cite{Viswanathan_Hussein_2017}. Depth cameras are commonly used with an RGB camera for object recognition, but are not generally used alone \cite{Ophoff_VanBeeck_Goedeme_2019}. While we only utilize one of the sensors to feed our gesture recognition algorithm, understanding the SDC sensor suite is crucial, as we encourage future experiments to incorporate data from other sensors as well. 

\subsubsection{Image Streamer}

Frames from the RGB camera of the SDC are streamed in real-time to a Python client. We observed an inverse relationship between the quality of the spatial dimensions (i.e. image resolution and angular size of the field of view) and temporal dimensions (i.e. sampling rate or Frames Per Second (FPS)) of the frames streamed. Higher spatial quality allows our algorithm to view the hand/arm in greater detail in each frame. Higher temporal quality allows our algorithm to view the entire motion of the gesture in greater detail. 

See Table~\ref{fig:ConfigFPS} for an illustration of this trade-off. Both of the resolutions we examine (\texttt{2840x2400} and \texttt{1280x1100}) are typical resolutions used by many autonomous vehicles, ranging from autonomous cars to golf carts etc. The low FPS shown in Table~\ref{fig:ConfigFPS} is simply an artifact of our experimental setup. Real vehicles process the data anywhere between 15 to 30 FPS, depending upon the speed of the vehicle. In our case, the car is always assumed to be halted before it analyzes a gesture and decides to move, such as at a stop sign or a crosswalk. A frame rate close to 15 FPS is sufficient for gesture recognition.

\begin{table}[b]
\small
\centering
\begin{tabular}{|l|l|l|l|}
\hline
\textbf{Width (px)} & \textbf{Height (px)} & \textbf{Field of View ($^\circ$)} & \textbf{Speed (FPS)} \\ \hline
2840       & 2400        & 90                      & 2.09        \\ \hline
1280       & 1100        & 40                      & 7.75        \\ \hline
1280       & 480        & 50                      & 12.62        \\ \hline
\end{tabular}
\caption{Increasing the information in the frames used by the SDC decreases the speed it can be streamed. We use the configuration with the highest FPS.} \label{fig:ConfigFPS}
\end{table}

To achieve the necessary frame rate, we optimize the system to be more balanced in three ways:
\begin{itemize}
\item (A) We decrease the image field of view from the default 90 degrees to 50 degrees. This enables us to decrease the width of the image without losing pixel density (DPI).
\item (B) We set the width and height of the streamed frames to \texttt{1280x480~px}.
\item (C) We set the clock speed of the simulator to 0.14 (1 second in reality : 0.14 seconds in the simulation).
\end{itemize}

(A) and (B) allow the image quality to be reduced but still usable by an algorithm. (C) gives the computer more time to get frames per every game second of the simulator. With this configuration, we pull one frame every 0.566 seconds, or 12.62 frames per simulator second.

Within the next section, we discuss the hand gesture recognition algorithm, a core part of the \gladas system. We will detail the dataflow and structure of the deep learning-based algorithm, which differs from traditional approaches.

\section{Algorithm Methodology}
\label{sec:algorithm}

\subsection{Overview}

\begin{figure*}
\centering
  \includegraphics[width=\textwidth]{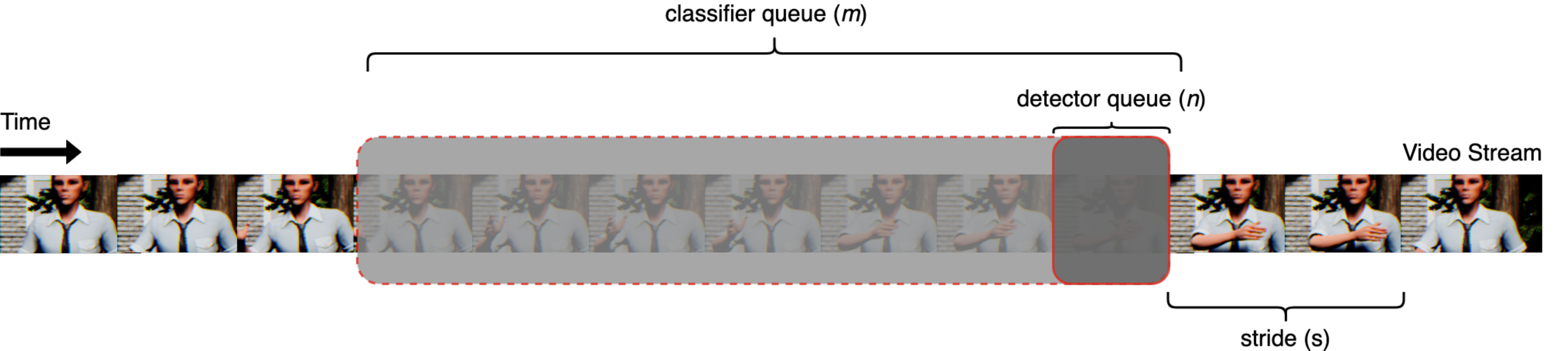}
  \caption{The general structure of the proposed two-model architecture. As frames are streamed in real-time, the lightweight detector processes every $s$-th frame, where $s$ is the stride. The detector, which identifies pedestrians within the SDC's field of vision, acts as a trigger for the more time-intensive gesture classifier, which operates on the previous 40 frames to classify the pedestrian's hand gesture.}~\label{fig:AlgorithmWorkflow}
\end{figure*}

\gladas is designed to be compatible with most gesture recognition algorithms that might be used with SDCs. We develop and demonstrate the testing of one such algorithm, which takes in RGB camera data (without image segmentation). Our algorithm incorporates two different models, cascaded sequentially. The first is used as a lightweight detector. \gladas first looks for a pedestrian, which is implemented as a Pedestrian Detector (PD). The PD model acts as a switch for the more complex classifier, the Gesture Classifier (GC). The GC is responsible for identifying the pedestrian's actual gesture.

Our real-time gesture recognition algorithm uses a sliding window approach, as illustrated in Figure~\ref{fig:AlgorithmWorkflow}. The PD (which receives an input of $n=1$ frame) processes the video stream with a stride (denoted as $s$) of five; it only operates on every fifth frame. A lower stride would allow us to detect pedestrians in more frames, but at the cost of more resources and computing power. If the PD positively detects a pedestrian, the GC is activated on the previous $m=40$ frames, detecting a gesture of one of the five classes: ``Stop'', ``Go Left'', ``Go Right'', ``Go Straight'', and ``No Gesture''. We elaborate on the architecture and image transformation process for each model below.

\subsection{Pedestrian Detector (PD)}

The detector (A) detects if a pedestrian exists, and (B) returns the coordinates of his/her upper body. The detector is designed to be as lightweight and fast as possible, so it can be applied in real-time to frames without heavy computational costs. 

\subsubsection{Pre-Processing}

\begin{figure}%
\centering
\subfigure[An example image streamed from the self-driving car's RGB camera.]{%
\label{fig:PD1}
\includegraphics[height=0.69in, width=0.4\columnwidth]{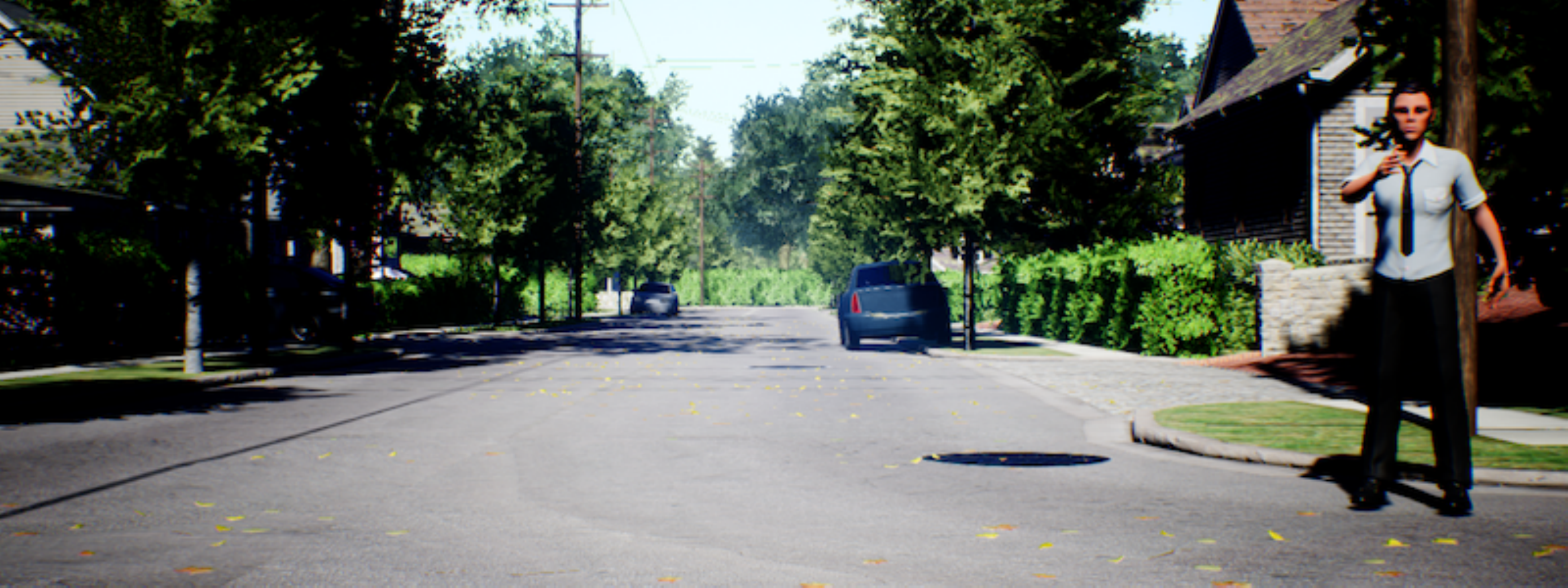}}
\qquad
\subfigure[The PD processes and crops the image to focus on the upper body.]{%
\label{fig:PD2}
\includegraphics[height=0.69in, width=0.4\columnwidth]{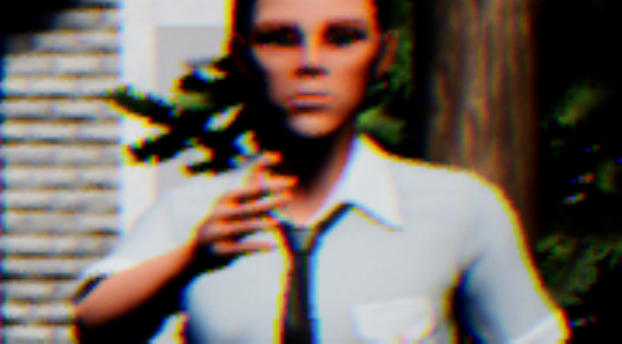}}%
\caption{Effect of the PD on the SDC's camera frames.}
\end{figure}

We scale the original \texttt{1280x480~px} RGB frame shown in ~\ref{fig:PD1} from the video stream to \texttt{768x288~px}, reducing the time requirement of the PD. These new width-height dimensions are 60\% of the original frame dimensions. A smaller input size reduces the processing time of the model.

\subsubsection{Model}
Our algorithm incorporates OpenCV's Pedestrian Recognition model with the Histogram of Oriented Gradients (HOGS)~\cite{Dalal_Triggs_2005} method. We input the smaller frame to the  model, which returns coordinates of a bounding box outlining the figure of any detected pedestrians.

\subsubsection{Post-Processing}

The bounding box coordinates, meant for the \texttt{768x288~px} image, are scaled up to fit the original \texttt{1280x480~px} image. The gesture classifier model is trained on image of the upper body, meaning we do additional cropping---1/7 of the top, 1/3 of the bottom, 1/9 of the left, and 1/5 of the right sides of the bounding box are removed. The final cropped coordinates, when applied to the car's original frame, result in an image similar to ~\ref{fig:PD2}. These new bounding box coordinates are sent to the Gesture Classifier model.

\subsection{Gesture Classifier (GC)}

The GC, if given a bounding box denoting a pedestrian, returns the gesture the pedestrian is making. This involves pre-processing the image, running a deep learning model, and post-processing the data in real-time for the car.

\subsubsection{Pre-Processing}

The classifier receives the prior 40 frames from the image streamer, as well as the
upper-body bounding box coordinates from the detector. Each of the 40 frames are cropped with these coordinates. We choose 40 frames, as it takes $\sim$40 frames to capture the entire motion of the pedestrian's gesture within the simulator. Next, we perform two transformations:

\begin{itemize}
    \item {\bf Temporal Transform.} Out of the 40 frames, we take a random sample of 32 frames in consecutive order, as the model requires only 32 frames for input.
    \item {\bf Spatial Transform.} We resize the 32 image frames to \texttt{112x112~px}, in order to meet the model's input requirements. The images are kept in RGB format.
\end{itemize}

\subsubsection{Model}

The GC incorporates a gesture recognition model by K\"op\"ukl\"u et al. \cite{Kopuklu_Gunduz_Kose_Rigoll_2019}. It is a 3D Convolutional Neural Network (3D CNN) that is pretrained on the 20BN-Jester Dataset~\cite{20bnjester}. We employ a 3D CNN due to its ability to process entire videos (i.e., groups of image frames organized by time). This is necessary in order to preserve the temporal semantics of the action. For example, the entire motion of a hand waving may signify ``Hello'', while taking just one still frame of that motion when instead it might in fact signify ``Stop'' over a series of static images.

The 20BN-Jester Dataset comprises 148,092 videos of humans enacting hand gestures, organized into 27 different categories. These categories include our 5 chosen hand gesture classes, as well as 22 other gestures with no immediate relevancy to pedestrian-to-car communication (e.g. Drumming Fingers). Each video generally features the chest to head of the human. 

The 32 frames of dimensions \texttt{112x112 px} are input to the model as a tensor. The model outputs 27 different class predictions, each with a corresponding confidence. We remove the 22 extraneous classes, focusing on the 5  gestures (``Go Forward'', ``Stop'', ``Go Right'', ``Go Left'', and ``No Gesture'') we selected. The gesture with the highest confidence is returned as the predicted gesture.


\section{Experimental Setup}
\label{sec:expsetup}

In order to test our hand gesture recognition algorithm, we set up several scenarios in \gladas. These scenarios are models of common Car-Pedestrian Interaction scenarios in real-life \cite{Chrysler_Ahmad_Schwarz_2015}, which best allow understanding of the algorithm's performance in typical, everyday driving. 

\begin{figure}%
\centering
\subfigure[Scenario One.]{%
\label{fig:Scen1}
\includegraphics[width=0.93\columnwidth]{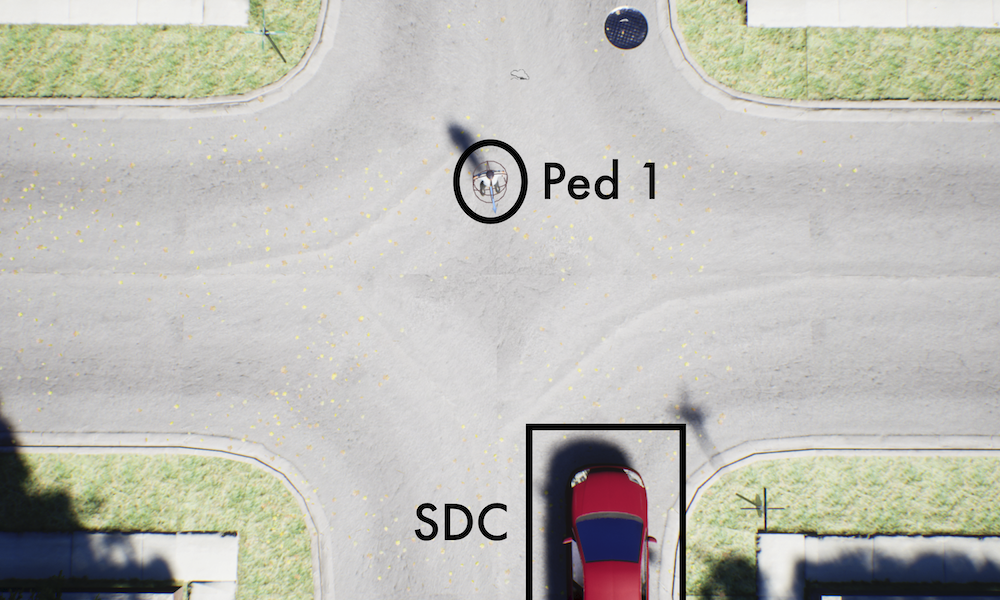}}%
\vspace*{2pt}
\subfigure[Scenario Two.]{%
\label{fig:Scen2}
\includegraphics[width=0.93\columnwidth]{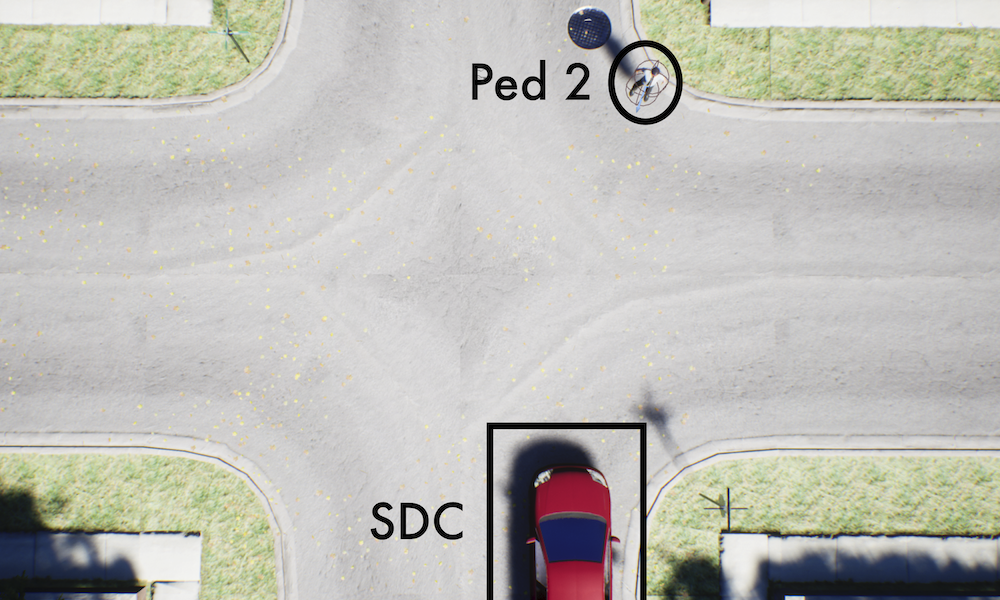}}%
\vspace*{2pt}
\subfigure[Scenario Three.]{%
\label{fig:Scen3}
\includegraphics[width=0.93\columnwidth]{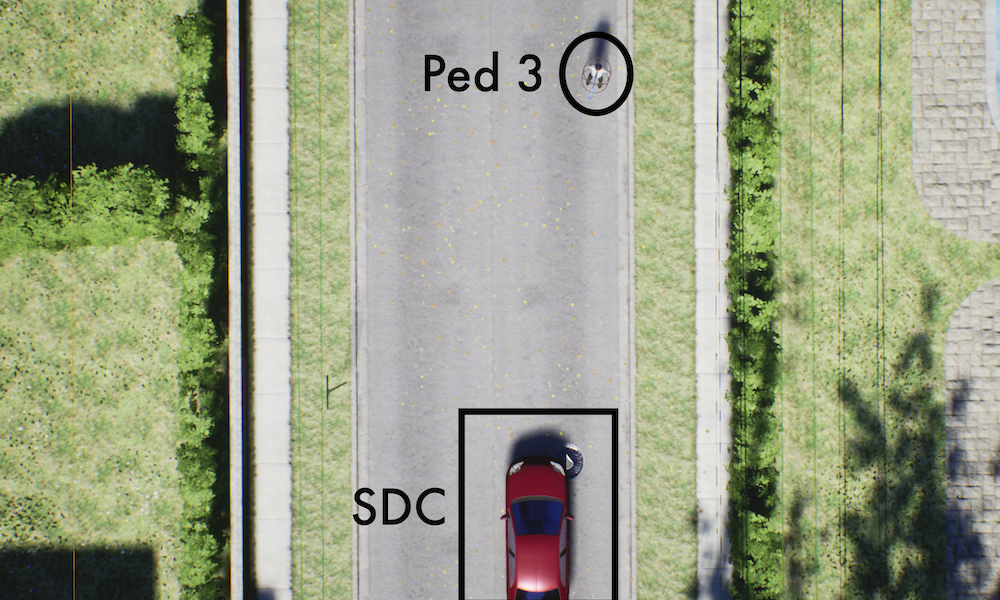}}%
\vspace*{2pt}
\subfigure[Scenario Four.]{%
\label{fig:Scen4}
\includegraphics[width=0.93\columnwidth]{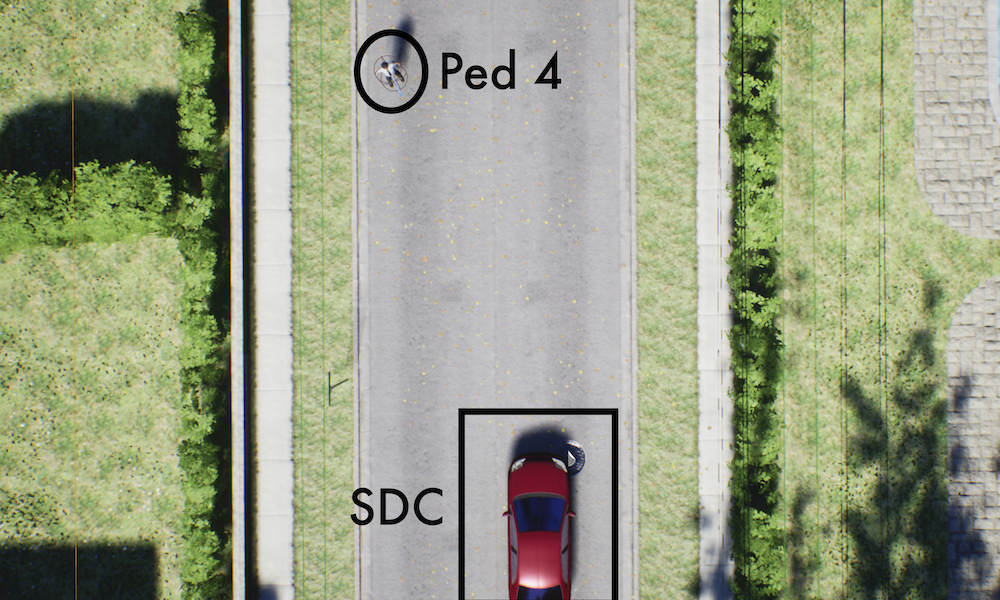}}%
\caption{The SDC and pedestrian are positioned as shown for the four CPI Scenarios.}
\end{figure}

\begin{enumerate}
\item Police Officer-Controlled 4-Way Intersection (\ref{fig:Scen1})
\item Pedestrian Crossing at 4-Way Intersection (\ref{fig:Scen2})
\item Pedestrian Crossing from Left at Mid-Block (\ref{fig:Scen3})
\item Pedestrian Crossing from Right at Mid-Block (\ref{fig:Scen4})
\end{enumerate}


We conduct tests of each of the four scenarios. Pedestrians only make relevant gestures during each scenario. In Scenario 1, the police officer uses ``Go Forward'', ``Stop'', ``Go Right'', ``Go Left'', and ``No Gesture''. In Scenarios 2-4, the pedestrian uses ``Go Forward'', ``Stop'', and ``No Gesture''. We define a scenario-gesture (SG) as the unique pairing of a gesture and scenario, for a total of 14 SG's. The algorithm is tested in each SG 2,000 times, for a total of 28,000 tests---a reasonable amount of iterations for understanding performance while not taking an inordinate amount of time to run. In these test scenarios, both the car and pedestrian are kept in the same position throughout.

We run our tests with an NVIDIA GeForce RTX 2080 TI Graphics Card, Intel Core i9-9940X CPU @ 3.30 GHz, and 31.7 Gigabytes of Random-Access Memory. Our results are detailed in the next section.

\section{Results}
\label{sec:results}

\subsection{Precision-Recall Analysis}

To understand our gesture recognition algorithm's performance in \gladas, we chose to analyze the GC's precision-recall characteristics as well as its accuracy---this analysis provides a good picture of our classifier's performance. In order to do this, we computed the classifier's confusion matrix---True Positive (TP), False Positive (FP), True Negative (TN), and False Negative (FN) metrics---for each gesture of each scenario over several classification-confidence thresholds.

For any given SG, TPs and FNs are obtained from the data produced by the 2,000 tests of that particular SG---ie. how often did the classifier properly recognize the gesture being shown. TNs and FPs are derived from the number of occurrences of the gesture in the other SG's associated with the same scenario---ie. how often did the classifier think it recognized a particular gesture even though that was not the gesture being shown. 

We make use of the following metrics in our analysis:
$$Precision = \frac{TP}{TP+FP}$$
$$Recall = \frac{TP}{TP+FN}$$ 

Importantly, we collect both the F1 score and accuracy of our classifier over the different SG combinations. While accuracy is a more intuitive metric, F1 score is a better measure of a classifier's performance in this context. Accuracy tells us only how often our classifier correctly identifies TPs and TNs, while F1 score additionally penalizes FNs and FPs---in the context of an SDC, these sorts of mis-classifications could lead to very negative outcomes.

$$Accuracy = \frac{TP + TN}{TP + TN + FP + FN}$$
$$F1\text{ }Score = 2 \times {\frac{Precision \times Recall}{Precision+Recall}}$$

Each precision-recall curve is generated by sweeping 1000 equally-spaced classification-confidence thresholds ($\delta$) between 0 and 1.0, computing the confusion matrix of each SG. The resulting precision-recall curves provide an insight into the trade-off between making our classifier more sensitive (ie. increasing recall) and making our classifier better at discriminating true positives (ie. increasing precision). 


\begin{figure}
    \centering
    \subfigure[Scenario One.]{
    \label{fig:Sc1}
    \includegraphics[width=0.95\columnwidth]{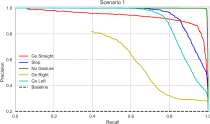}}
\qquad
    \centering
    \subfigure[Scenario Two.]{
    \label{fig:Sc2}
    \includegraphics[width=0.95\columnwidth]{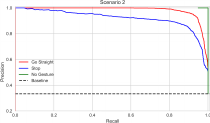}}
\qquad
    \centering
    \subfigure[Scenario Three.]{
    \label{fig:Sc3}
    \includegraphics[width=0.95\columnwidth]{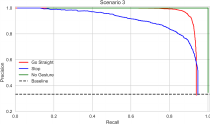}}
\qquad
    \centering
    \subfigure[Scenario Four.]{
    \label{fig:Sc4}
    \includegraphics[width=0.95\columnwidth]{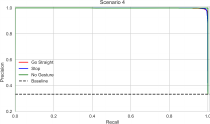}}
    \caption{Precision-Recall curves of the classifier for different scenario-gesture pairs.}~\label{fig:PrecisionRecallResults}
\end{figure}

Our precision-recall curves are shown in Figure~\ref{fig:PrecisionRecallResults}. Inspection of these precision-recall curves suggests the following about our gesture recognition algorithm's GC. First, recognition of gestures across all four scenarios is better than the baseline, which is equivalent to a random guess. Second, recognition of ``No Gesture'' across all scenarios is almost perfect---most likely, the GC has a particularly easy time discriminating that particular ``gesture'' within the captured video of the pedestrian. Third, recognition of ``Go Right'' in Scenario One is relatively poor when compared to the rest of the gestures---one potential explanation for this is that the GC may have regularly mis-identified ``Go Right'' gestures as ``Go Left'' gestures due to their similar hand motions. Finally, the precision-recall curves for Scenario Four suggest that the GC is performing virtually identical to an ``ideal'' classifier in this scenario. While the precision-recall curves suggest the GC has fairly decent classification performance over most SGs, the performance implied by the precision-recall curves from Scenario Four does appear to be a bit of an outlier. In particular, Scenarios One and Four are similar with regards to the relative positions of the car and pedestrian---different lighting angles and backgrounds behind the pedestrians may have affected the disparity in classification performance between the scenarios.

\begin{table}[]
\small
\centering
\begin{tabular}{@{}ccccccc@{}}
\cmidrule(l){3-7}
\multicolumn{2}{l}{\multirow{2}{*}{}}                                          & \multicolumn{5}{c}{\textbf{Gesture}}                                                                                                                                                                                                               \\ \cmidrule(l){3-7} 
\multicolumn{2}{l}{}                                                           & \multicolumn{1}{l}{\textit{\textbf{Go Straight}}} & \multicolumn{1}{l}{\textit{\textbf{Stop}}} & \multicolumn{1}{l}{\textit{\textbf{No Gesture}}} & \multicolumn{1}{l}{\textit{\textbf{Go Right}}} & \multicolumn{1}{l}{\textit{\textbf{Go Left}}} \\
\multicolumn{1}{|c|}{\multirow{4}{*}{\rot{\textbf{Scenario}}}} & \textit{\textbf{1}} & 93.4\%                                            & 95.6\%                                     & 93.3\%                                           & 88.2\%                                         & 95.0\%                                        \\
\multicolumn{1}{|c|}{}                                   & \textit{\textbf{2}} & 94.9\%                                            & 93.3\%                                     & 98.1\%                                           & n/a                                            & n/a                                           \\
\multicolumn{1}{|c|}{}                                   & \textit{\textbf{3}} & 93.6\%                                            & 91.2\%                                     & 96.4\%                                           & n/a                                            & n/a                                           \\
\multicolumn{1}{|c|}{}                                   & \textit{\textbf{4}} & 92.3\%                                            & 98.7\%                                     & 99.9\%                                           & n/a                                            & n/a                                          
\end{tabular}
\caption{The accuracy of each Scene-Gesture pair.} \label{fig:acctable}

\vspace*{0.5 cm}

\small
\centering
\begin{tabular}{@{}ccccccc@{}}
\cmidrule(l){3-7}
\multicolumn{2}{l}{\multirow{2}{*}{}}                                          & \multicolumn{5}{c}{\textbf{Gesture}}                                                                                                                                                                                                               \\ \cmidrule(l){3-7} 
\multicolumn{2}{l}{}                                                           & \multicolumn{1}{l}{\textit{\textbf{Go Straight}}} & \multicolumn{1}{l}{\textit{\textbf{Stop}}} & \multicolumn{1}{l}{\textit{\textbf{No Gesture}}} & \multicolumn{1}{l}{\textit{\textbf{Go Right}}} & \multicolumn{1}{l}{\textit{\textbf{Go Left}}} \\
\multicolumn{1}{|c|}{\multirow{4}{*}{\rot{\textbf{Scenario}}}} & \textit{\textbf{1}} & 76.7\%                                            & 84.9\%                                     & 83.2\%                                           & 62.6\%                                         & 83.1\%                                        \\
\multicolumn{1}{|c|}{}                                   & \textit{\textbf{2}} & 88.8\%                                            & 87.0\%                                     & 96.4\%                                           & n/a                                            & n/a                                           \\
\multicolumn{1}{|c|}{}                                   & \textit{\textbf{3}} & 85.4\%                                            & 82.2\%                                     & 93.3\%                                           & n/a                                            & n/a                                           \\
\multicolumn{1}{|c|}{}                                   & \textit{\textbf{4}} & 81.9\%                                            & 97.5\%                                     & 99.7\%                                           & n/a                                            & n/a                                          
\end{tabular}
\caption{The F1 Score of each Scene-Gesture pair.} \label{fig:f1table}
\end{table}

In order to measure the accuracy and F1 score of the GC's performance over the four scenarios, we choose a particular classification-confidence threshold, $\delta=0.40$, which provides a good trade-off between precision and recall. The GC's accuracy and F1 score over all scenario and gesture combinations tested are shown in Table~\ref{fig:acctable} and Table~\ref{fig:f1table}, respectively. These results suggest an average F1 score of 85.91\% for the GC and an average accuracy of 94.56\%. That the classifier accuracy is larger than the classifier F1 score suggests the presence of a significant number of FPs and FNs---considering only accuracy would provide an overly optimistic perception of the classifier's performance. The F1 score suggests that the GC would \textit{not} identify the correct gesture about 14.09\% of the time. In the context of roadway driving, minimizing this rate and maximizing classification accuracy is crucial, as each classification error could result in the car making an incorrect move---driving forward after incorrectly thinking a pedestrian's ``Stop'' gesture was the ``Come'' gesture. These mistakes could result in serious collisions with pedestrians, policepeople, and other cars. 

\subsection{Implications}

The main goal of our research was to understand the ability of self-driving cars to understand human hand gestures in CPI scenarios. A realistic simulator with multiple scenarios requiring Pedestrian-to-Car communication was developed as part of the \gladas system. Hand gestures were performed by each pedestrian, allowing \gladas to support testing of the SDC's hand gesture recognition algorithm. 

\subsubsection{Research Needs}
In order for us to release reliable SDC hand gesture recognition algorithms that save, not harm lives, they must be able to reliably detect, classify, and react to hand gestures virtually perfectly. Our results show that self-driving cars classify pedestrian hand gestures with an F1 score of roughly 85.91\% with our baseline algorithm, a phenomenon that leads to potential crashes with cars, pedestrians, and policemen 14.09\% of the time. These results strongly enforce the need for continued research and development of better algorithms, or at the very least, deep consideration of the fail-safes that need to be built into an SDC system to safely recover from gesture classification errors. 

Due to the intrinsic plug-and-play nature of \gladas, a host of subsequent studies can be supported in the testing of alternative methods, comparing their results to ours. Future algorithms could incorporate additional processing, such as image segmentation to improve recognition capabilities. 

\subsubsection{Benchmarking}
Additionally, our methodology serves as a foundation for future GL benchmarking efforts. As an analog to current seat belt and crash safety tests---standardized by the NHTSA \cite{nhtsa_2019} to ensure roadway safety---as well as human driver tests meant to ensure minimal levels of driving competency, SDCs will have to be heavily tested for reliability, accuracy, and safety. 

Gesture learning tests, as one component in a greater SDC benchmarking suite, are therefore necessary. \gladas illustrates one method of doing so, scrutinizing the performance of gesture recognition models in a simulated environment within the context of important driving situations. \gladas' portability to other models makes it a good baseline for development of future GL benchmarking tests. 

\subsubsection{Bias}
Limiting dataset bias with regards to factors such as regional hand gesture signals, skin tones, and worn garments is important for a practical, real-world application of gesture learning---different areas around the world may use different hand gestures, feature different skin tones, and clothing may obfuscate hand gestures (eg. gloves worn during cold weather). 

We recommend the use of \gladas as a platform to integrate and account for such regional differences in future work.

\section{Conclusion}
\label{sec:conc}

We presented \gladas, a simulator-based framework for research on human-AV interaction. We utilize \gladas to test a hand gesture recognition algorithm, developed using two pre-trained models: a pedestrian detector, and a hand gesture classifier. We challenge the algorithm to recognize a simulated pedestrian's hand gesture in four common Car-Pedestrian interaction scenarios, using \gladas to evaluate its effectiveness. The results provided by the simulator suggest a need for continued gesture learning research as well as the necessity of developing benchmarking and safety tests for self-driving cars, particularly within the context of gesture learning. We hope \gladas  inspires and enables further research into self-driving car hand gesture recognition, paving the way for full autonomy. For instance, future work could involve using more advanced gesture recognition solutions that are based on instance segmentation to more clearly identify the gestures.

\section{Acknowledgments}
We would like to express our appreciation to the Research Science Institute (RSI) and Center for Excellence in Education (CEE) for putting the authors of this paper in touch. 

%
%
%
%
%

\balance{}

\bibliographystyle{SIGCHI-Reference-Format}
\bibliography{sample}

\end{document}